\documentclass[12pt]{iopart}

\usepackage{iopams,graphicx,amssymb}
\eqnobysec

\begin{document}

\title[Scaling in erosion of landscapes]
{Scaling in  erosion of landscapes:
Renormalization group analysis of a model with infinetly many couplings}

\author{N. V. Antonov and P. I. Kakin}

\address{Department of Theoretical Physics, St.~Petersburg State University,
Uljanovskaja 1, Petrodvorez, St.~Petersburg, 198504 Russia}

\ead{n.antonov@spbu.ru, p.kakin@spbu.ru}

\begin{abstract}
Standard field theoretic renormalization group  is applied to the model of landscape erosion introduced 
by R. Pastor-Satorras and D. H. Rothman [{\it Phys. Rev. Lett.}  {\bf 80}: 4349 (1998); 
{\it J. Stat. Phys.} {\bf 93}: 477 (1998)] yielding unexpected results: the 
model is multiplicatively renormalizable only 
if it involves infinitely many coupling constants, ({\it i.e.}, the corresponding renormalization 
group equations involve 
infinitely many $\beta$-functions). Despite this fact, the one-loop counterterm
can be derived albeit in a closed form in terms of the certain function $V(h)$, entering the original 
stochastic equation, 
and its derivatives with respect to the height field $h$. Its Taylor expansion gives rise to the full infinite
set of the one-loop renormalization constants, $\beta$-functions and anomalous dimensions. 
Instead of a set of fixed points, there is a two-dimensional surface of fixed points that is likely 
to contain infrared attractive region(s). If that is the case, the model exhibits scaling 
behaviour in the infrared range. The
corresponding critical exponents are nonuniversal through the dependence
on the coordinates of the fixed point on the surface, but satisfy certain universal exact relations.

\end{abstract}


\section{Introduction and description of the model} \label{sec:Intro}

Over decades, constant interest has been attracted to the problem of landscape
erosion due to the flow of air or water over it, and to related
problems like, {\it e.g.} granular flows; see Refs. \cite{E1}--\cite{E14}
and the literature cited therein.
Of course, those issues concern a wide variety of diverse physical phenomena;
the underlying dynamical models have been a source of much controversy
\cite{E5}--\cite{E13}.
However, in analogy with critical phenomena, one can hope that universal
aspects of landscape erosion (like the exponents in scaling laws) can be
described within the framework of relatively simple semiphenomenological
models, constructed on the basis of dimensionality and symmetry
considerations; see, {\it e.g.} the discussion in \cite{Pastor1,Pastor2} and
references therein.

Similar situation takes place in the related problem of kinetic
roughening of surfaces or interfaces, described by the well known
Kardar-Parisi-Zhang stochastic model \cite{KPZ}
and its descendants \cite{Kim}--\cite{AK1}. Another example is provided
by the problem of self-organized criticality, which in the continuum
limit is described by the Hwa-Kardar stochastic model \cite{HK}
and its modifications \cite{Tadic,AK2}.

For the erosion of a surface with a fixed mean tilt, analogous model was
proposed in \cite{Pastor1,Pastor2}. Let us describe that model first.

Let ${\bf n}$ be a unit constant vector that determines a certain preferred
direction (direction of the slope) and, therefore,
introduces intrinsic anisotropy into the model.
Then any vector can be decomposed into the components perpendicular and
parallel to ${\bf n}$. In particular, for the $d$-dimensional
horizontal position ${\bf x}$ one has
${\bf x} = {\bf x}_{\bot} + {\bf n} x_{\parallel}$ with
${\bf x}_{\bot} \cdot {\bf n} =0$.
In the following, we denote the derivative in the full $d$-dimensional
${\bf x}$ space by
$\partial = \partial/ \partial {x_i}$ with $i=1\dots d$, and the
derivative in the subspace orthogonal to ${\bf n}$ by
$\partial_{\bot}=\partial/ \partial {x_{\bot i}}$ with $i=1\dots d-1$.
Then the derivative in the parallel direction is written as 
$\partial_{\parallel} = {\bf n}  \cdot \partial$.

The stochastic differential equation for the height of the profile,
{\it i.e.} for the height field $h(x)=h(t,{\bf x})$, proposed in
\cite{Pastor1,Pastor2} is taken in the form
\begin{equation}
\partial_{t} h= \nu_{\bot}\, \partial_{\bot}^{2} h + \nu_{\parallel}\,
\partial_{\parallel}^{2} h +
\partial_{\parallel}^{2} V(h) + f.
\label{eq1}
\end{equation}
Here $\partial_{t} = \partial/ \partial {t}$, $\nu_{\parallel}$ and
$\nu_{\bot}$ are topographic diffusion coefficients, $V(h)$ is some
function that depends only on the field $h(x)$ (and not on its derivatives)
and $f(x)$ is a Gaussian random noise with zero mean and prescribed
pair correlation function
\begin{equation}
\langle f(x)f(x') \rangle = D
\delta(t-t')\, \delta^{(d)}({\bf x}-{\bf x}')
\label{forceD}
\end{equation}
with some positive amplitude $D$. Detailed discussion of the derivation
of the model (\ref{eq1}), (\ref{forceD})
and its relationship to other models of erosion and
self-organized criticality is given in \cite{Pastor1,Pastor2}.

The function $V(h)$ understood as series in powers of $h$. In \cite{Pastor1,Pastor2}
is was taken odd in $h$: this is dictated by the
symmetry $h,f\to-h,-f$; another symmetry of the model is
$x_{\parallel}\to-x_{\parallel}$. The authors of \cite{Pastor1,Pastor2}
truncated the Taylor expansion of $V(h)$ on the leading $h^{3}$ term
(the term linear in $h$ is written in (\ref{eq1}) separately) and then
applied to the resulting model the dynamic Wilsonian renormalization group (RG)
and the expansion in $4-d$, the deviation of the dimension $d$
from its supposed upper critical value $d=4$. In the leading one-loop order, they
established existence of the infrared (IR) attractive fixed point and
calculated the corresponding critical (roughness) exponents in a good
agreement with the experimental data obtained from sea floor measurements.

In the present paper we apply to the model \cite{Pastor1,Pastor2} the standard field
theoretic RG and arrived at completely different results.
The plan of the paper and the main results are as follows.

In section \ref{sec:Model} we present the field theoretic formulation of
the stochastic problem (\ref{eq1}), (\ref{forceD}) for the arbitrary
(not necessarily odd) full-scale (not truncated) function $V(h)$.

In section \ref{sec:Reno} we discuss ultraviolet (UV) divergences and
renormalization procedure of the resulting field theory. We show that the 
upper critical dimension is in fact $d=2$. This leads to drastic change in the 
RG analysis of the model. Namely, 
the higher-order terms of the Taylor expansion of
$V(h)$ cannot be dropped, because they unavoidably appear as counterterms
in the correct renormalization procedure. In other words, any truncated
model is not multiplicatively renormalizable. This means that the properly
constructed renormalized model necessarily involves infinitely many coupling
constants, and the corresponding RG equations involve infinitely many
$\beta$-functions. This also means that the RG analysis performed in
Refs.~\cite{Pastor1,Pastor2} for the truncated model is not self-consistent
and its results cannot be considered reliable.

We write down
the corresponding renormalized action functional, renormalization
relations for the fields and parameters, RG equations and RG functions
($\beta$-functions and anomalous dimensions).

In section~\ref{sec:Funk} we explicitly perform the renormalization
in the leading one-loop order. The key point is that, despite the fact that the model
involves infinitely many couplings, the one-loop counterterm
can be derived in a closed form in terms of the function $V(h)$
and its derivatives. Its Taylor expansion gives rise to the full infinite
set of one-loop renormalization constants, and, therefore, to all
$\beta$-functions and anomalous dimensions.

In this derivation, we adopt the functional method applied earlier by
A.~N.~Vasil'ev and one of the authors \cite{AV} to an isotropic model
of surface roughening, proposed in \cite{Pavlik} as a possible
modification of the Kardar-Parisi-Zhang equation; see also \cite{AA,AAA}.

In section~\ref{sec:Att} we analyze attractors of the obtained RG
equations in the infinite-dimensional space of coupling constants.
It turns out, that instead of a set of fixed points (like for most
multicoupling models), there is a two-dimensional surface of fixed points.
For odd $V(h)$, that is, for the model \cite{Pastor1,Pastor2}, it
reduces to a curve. It seems likely that it contains IR attractive region(s).
If so, the model exhibits scaling behaviour in the IR range. The
corresponding critical exponents are nonuniversal through the dependence
of the coordinates of the fixed point on the surface (curve), but
satisfy certain exact relations.

Possible consequences for the comparison with the experiments and
remaining problems are briefly discussed in section~\ref{sec:Conc}.

\section{Field Theoretic Formulation of the Model} \label{sec:Model}

According to the general statement (see, {\it e.g.} the books \cite{Zinn,Book3}
and the references therein), 
the stochastic problem
(\ref{eq1}), (\ref{forceD}) is equivalent to the field theoretic model
of the doubled set of fields $\Phi=\{h,h'\}$ with the action functional  
\begin{equation}
{\cal S}(\Phi)=h'h'+h'\left\{-\partial_{t}h+\nu_{0\bot}\, \partial_{\bot 0}^{2} h + \nu_{\parallel 0}\,
\partial_{\parallel}^{2} h +
\partial_{\parallel}^{2}\sum_{n=2}^{\infty}\frac{\lambda_{n0} h^{n}}{n!}\right\}
\label{act1}
\end{equation}
(we have scaled out $D_0$ and other factors of $h'h'$ by adjusting the values of $\lambda_{n0} $).
Here and below, all the needed integrations over $x = (t,{\bf x})$ and summations
over repeated tensor indices are always implied, {\it e.g.}
\begin{equation}
h' h'=\int dt\int d{\bf x} \,\, h'(t,{\bf x})\, h'(t,{\bf x}).
\end{equation}
The subscript 0 means that the parameters in (\ref{act1}) are not yet renormalized (bare).

The field theoretic formulation means that various correlation and
response functions of the stochastic problem (\ref{eq1}), (\ref{forceD})
can be identified with various Green's functions of the field theoretic model
with the action (\ref{act1}). In other words, they are represented by
functional averages over the full set of fields $\Phi=\{ h,h'\}$ with
the weight $\exp {\cal S}(\Phi)$.

\section{UV divergences and renormalization} \label{sec:Reno}

The  analysis of canonical dimensions is employed to analyze the UV divergences; see, {\it e.g.} \cite{Zinn,Book3}.
Conventional  dynamic models of the type (\ref{act1}) have two scales, and their dimensions are described by   
the  two numbers - the frequency dimension $d_{F}^{\omega}$, and the momentum dimension $d_{F}^{k}$. 
They completely define the canonical dimension of a quantity $F$ (a field or a parameter), 
 and  are determined so that $[F] \sim [T]^{-d_{F}^{\omega}} [L]^{-d_{F}^{k}}$, where $L$ is the typical length
scale and $T$ is the time scale; see, {\it e.g.} Chap.~5 in book \cite{Book3}.
In the present case, however, due to the anisotropy there are two independent momentum scales,
related to the directions perpendicular and parallel to the vector ${\bf n}$ which requires
a more detailed specification. Namely, two independent momentum canonical dimensions 
$d_{F}^{\bot}$ and $d_{F}^{\parallel}$ had to be introduced so that
\[ [F] \sim [T]^{-d_{F}^{\omega}}  [L_{\bot}]^{-d_{F}^{\bot}}
[L_{\parallel}]^{-d_{F}^{\parallel}}, \]
where $L_{\bot}$ and $L_{\parallel}$ are (independent) length scales in the
corresponding subspaces. The obvious
normalization conditions are $d_{k_{\bot}}^{\bot}= -d_{\bf x_{\bot}}^{\bot}=1$,
$d_{k_{\bot}}^{\parallel}=-d_{\bf x_{\bot}}^{\parallel}=0$,
$d_{k_{\bot}}^{\omega} = d_{k_{\parallel}}^{\omega}=0$,
$d_{\omega }^{\omega }=-d_t^{\omega }=1$, {\it etc.}; the
requirement that each term of the action functional (\ref{act1})
be dimensionless (with respect to all the three independent dimensions
separately) is the last condition needed to find the dimensions. The original momentum dimension can be found from the
relation $d_{F}^{k} = d_{F}^{\bot}+ d_{F}^{\parallel}$.
Then, based on $d_{F}^{k}$ and $d_{F}^{\omega}$, the
total canonical dimension can be introduced $d_{F}=d_{F}^{k}+2d_{F}^{\omega}  =
d_{F}^{\bot} + d_{F}^{\parallel} +2d_{F}^{\omega}$ (in the free theory,
$\partial_{t}\propto\partial^{2}_{\bot} \propto \partial^{2}_{\parallel}$),
which plays in the theory of renormalization of dynamic models the same
role as the conventional (momentum) dimension does in static problems;
see, {\it e.g.} Chap.~5 in book \cite{Book3}.

The canonical dimensions for the model (\ref{act1}) are presented in
table~\ref{table1}. The renormalized parameters (without the subscript 0) and the renormalization mass
$\mu$ will be introduced later.

\begin{table}[h]
\centering
\caption{Canonical dimensions of the fields and the parameters in the
model (\protect\ref{act1})} \label{table1}
\begin{tabular}{lllllllll}
\hline
$F$ & $h'$ & $h$ & $\nu_{\bot}$ & $\nu_{\parallel}$ & $\lambda_{n0}$ & $g_{n0}$ & $g_n$ & $\mu$\\\hline
$d_{F}^{\omega}$ & $1/2$ & $-1/2$ & $1$ & $1$ & $(n+1)/2$ & $0$ & $0$ & $0$\\
$d_{F}^{\parallel}$ & $1/2$ & $1/2$ & $0$ & $-2$ & $-(n+3)/2$ & $0$ & $0$&  $0$\\
$d_{F}^{\bot}$ & $(d-1)/2$ & $(d-1)/2$ & $-2$ & $0$ & $(d-1)(1-n)/2$ & 
$(2-d) (n-1)/2$ & $0$ & $1$\\
\hline
$d_{F}$ & $d/2+1$ & $-(2-d)/2$ & $0$ & $0$ & $(2-d)(n-1)/2$ 
& $(2-d) (n-1)/2$ & $0$ & $1$\\
\hline
\end{tabular}
\end{table}

From table~\ref{table1} we see that all the coupling constants  $g_{n0}$ become simultaneously dimensionless 
at $d=2$. This means that $d=2$ is the upper critical dimension for the full-scale model.
For this value of $d$, the total canonical dimension of the field $h$
vanishes. As explained below, this fact leads to serious consequences for the renormalization procedure.
This fact also means that  UV divergences in the Green's functions of the full-scale model
manifest themselves as poles in $\varepsilon=2-d$, and that $\varepsilon$
plays the role of the expansion parameter in the RG expansions.

The total canonical dimension of an arbitrary 1-irreducible Green's function
$\Gamma = \langle\Phi \cdots \Phi \rangle_{\rm 1-ir}$ with $\Phi=\{h,h'\}$
in the frequency--momentum representation is given by the relation:
\begin{equation}
d_{\Gamma}=d+2-d_h N_h-d_{h'}N_{h'},
\label{dGamma}
\end{equation}
where $N_h,N_{h'}$ are the numbers of the corresponding fields entering
into the function $\Gamma$; see, {\it e.g.} \cite{Book3}.

The total dimension $d_{\Gamma}$ in the logarithmic theory ({\it i.e.} at
$\varepsilon=0$) is, in fact, the formal index of the UV divergence:
$\delta_{\Gamma}=d_{\Gamma}|_{\varepsilon=0}$.
The superficial UV divergences, whose removal requires counterterms, can be
present only in those functions $\Gamma$ for which $\delta_{\Gamma}$ is
a non-negative integer. The counterterm is a polynomial in frequencies and
momenta of degree $\delta_{\Gamma}$ (given that the convention that
$\omega\propto k^2$ is implied).

If a number of external momenta occurs as an overall
factor in all diagrams of a certain Green's function, the real index of
divergence $\delta_{\Gamma}'$ will be smaller than $\delta_{\Gamma}$ by
the corresponding number. This is exactly what happens in our
model: using integration by parts, the derivative at the vertex 
$h'\partial_{\parallel}^2 V(h)$ can be moved onto the field  $h'$.
 This means that any appearance
of $h'$ in some function $\Gamma$ gives a square of such an external momentum, and the
real index of divergence is given by the expression
$\delta_{\Gamma}'= \delta_{\Gamma} - 2N_{h'}$. Moreover, $h'$ can appear
in the corresponding counterterm only in the form of derivative.

From table~\ref{table1} and the expression (\ref{dGamma}) one obtains:
\begin{equation}
\delta_{\Gamma}'= \delta_{\Gamma} - 2N_{h'}=4 - 4N_{h'} .
\label{IndeX}
\end{equation}

It is sufficient to consider only the case $N_{h'}>0$ because all the 1-irreducible Green's functions without the
response fields vanish identically in dynamical models (their diagrams always involve closed
circuits of retarded lines); see, {\it e.g.} \cite{Book3}. 

Straightforward analysis of the expression (\ref{IndeX}) shows that
superficial UV divergences can be present only in the
1-irreducible functions of the form $\langle  h'h\dots h \rangle_{1-ir} $ with 
the counter-term $(\partial_{\parallel}^2 h')h^n$ (for any $n\geq 1$). Indeed, all the other 
counter-terms ({\it e.g.} $h'h'$, $h'\partial_t h$, $h'\partial_{\bot}^2 h$) are not needed 
as the corresponding 1-irreducible functions are finite.

As all the terms $(\partial_{\parallel}^2 h')h^n$ are present
in the action (\ref{act1}), the full model is multiplicatively renormalizable.
The renormalized action  can be written in the form:
\begin{equation}
{\cal S}_R (\Phi)=h'h'+h'\left\{-\partial_{t}h+Z_{\bot}\nu_{\bot} 
\partial_{\bot}^{2} h + Z_{\parallel}\nu_{\parallel}\,
\partial_{\parallel}^{2} h +
\partial_{\parallel}^{2}\sum_{n=2}^{\infty}\frac{Z_n\lambda_{n} h^{n}}{n!}\right\}.
\label{RenAct}
\end{equation}

Here $\nu_{\bot}$, $\nu_{\parallel}$ and $\lambda_n$ are renormalized analogs of the bare
parameters (those with subscript 0).
The renormalization constants $Z_{\bot}$, $Z_{\parallel}$, and $Z_n$ depend only on the completely
dimensionless parameters $g_n$ and absorb the poles
in $\varepsilon$. The bare charges $g_{0}=\{g_{n0}\}$ and completely 
dimensionless renormalized charges $g=\{g_{n}\}$ ($n=2,3,\dots$)  are expressed in terms of 
bare parameters $\lambda_{n0}$ and renormalized parameters $\lambda_n$ as follows:
\begin{equation}
\lambda_{n0}=g_{n0} \nu_{\parallel 0}^{(n+3)/4}\nu_{\bot 0}^{(n-1)/4}, \quad
\lambda_{n}=g_{n} \nu_{\parallel}^{(n+3)/4}\nu_{\bot}^{(n-1)/4}\mu^{\varepsilon(n-1)/2},
\label{I6}
\end{equation}
Here the renormalization mass $\mu$ is an additional parameter of the renormalized theory; its canonical 
dimensions
are shown in table~\ref{table1}. 

The renormalized action (\ref{RenAct}) is obtained from the original
one (\ref{act1}) by the renormalization of the parametrs (the renormalization of the fields $h,h'$ is not required):
\begin{equation}
\nu_{\parallel 0} =\nu_{\parallel} Z_{\parallel}, \quad  \nu_{\bot 0} =\nu_{\bot} Z_{\bot}, \quad
g_{n0}=\mu^{\varepsilon(n-1)/2} g_{n}Z_{g_{n}}, \quad \lambda_{n0} = \lambda_n Z_n.
\label{I7}
\end{equation}

The renormalization constants in Eqs. (\ref{RenAct}) and (\ref{I7})
are related as follows:
 \begin{equation}
Z_{g_{n}}=Z_{n}Z_{\parallel}^{-(n+3)/4}Z_{\bot }^{-(n-1)/4}.
\label{I8}
\end{equation}

Let us consider an elementary derivation of the RG equations \cite{Zinn,Book3}.
The RG equations are written for the renormalized Green's functions
$G_{R} =\langle \Phi\cdots\Phi\rangle_{R}$. In the present case they are equal to
the original (unrenormalized) Green's functions $G$: $G(e_{0},\dots) = 
G_{R}(e,\mu,\dots)$ (because there is no renormalization for the fields) and, therefore, 
can be equally used for analyzing the critical behaviour. Here, 
$e_{0}=\{g_{n0}, \nu_{\parallel 0}, \nu_{\bot 0}, \dots \}$ is a full set of
bare parameters and $e=\{g_{n}, \nu_{\parallel}, \nu_{\bot}, \dots\}$ are their renormalized
counterparts; the ellipsis stands for the other arguments (times,
coordinates, momenta etc.).

We use $\widetilde{\cal D}_{\mu}$ to denote the differential operation
$\mu\partial_{\mu}|_{e_0}$. When expressed in the renormalized variables
it looks as follows:
\begin{equation}
{\cal D}_{RG}\equiv {\cal D}_{\mu} + \sum_{n=2}^{\infty}\beta_{n}\partial_{g_n}
- \sum_{F=\nu_{\parallel},\nu_{\bot}}\gamma_{F}{\cal D}_{F},
\label{RG2}
\end{equation}
where ${\cal D}_{x}\equiv x\partial_{x}$ for any variable
$x$. The anomalous dimensions $\gamma$ are defined as
\begin{equation}
\gamma_{F}\equiv \widetilde {\cal D}_{\mu} \ln Z_{F} \quad
{\rm for\ any\ quantity} \ F,
\label{RGF1}
\end{equation}
and the $\beta$ functions for the dimensionless coupling constants $g_n$ are
\begin{equation}
\beta_{n} \equiv \widetilde {\cal D}_{\mu} g_n = g_n\,[-\varepsilon(n-1)/2-\gamma_{g_n}].
\label{betagw}
\end{equation}

\section{One-loop expressions for the counterterm, renormalization
constants and RG functions} \label{sec:Funk}

Let us turn to the calculation of the constants $Z$ in the one-loop approximation. 
Despite the fact that the full renormalizable model involves infinitely many coupling constants, the one-loop
counterterm can be calculated in an explicit closed form in terms of the function $V(h)$. 

Consider the expansion of the generating functional $\Gamma_{R}(\Phi)$ of the 1-irreducible 
Green's's functions of our model in the number $p$ of loops:
    \begin{equation} 
    \Gamma_{R}(\Phi)=\sum_{p=0}^{\infty}  \Gamma^{(p)}(\Phi),\
    \Gamma^{(0)}(\Phi) = S_{R} (\Phi).
    \label{W19}
    \end{equation}
The loopless (tree-like) contribution is simply the action while the one-loop contribution can be calculated 
via following relation, see, {\it e.g.} \cite{Book1}:
    \begin{equation}
    \Gamma^{(1)}(\Phi) = - (1/2) {\rm Tr\ ln} (W/W_{0}),
    \label{W20}
    \end{equation}
where $W$ is a linear operation with the kernel
    \begin{equation}
    W(x,y)=-\delta^{2}S_{R} (\Phi) / \delta\Phi(x)\delta\Phi(y),
    \label{W21}
    \end{equation}
and $W_{0}$ is the similar expression for the free parts of the action. 
The both $W$ and $W_{0}$ are $2\times2$-matrices in the pair $ \Phi= \{h,h'\}$.

The requirement that UV divergences in (\ref{W19}) are removed, along with the minimal subtraction prescription,
 provides the uniquely determined values for constants $Z$. In the one-loop approximation 
we put $Z=1$ in (\ref{W20}) while keeping leading-order terms in the coupling constants 
$g_n$ in the loopless contribution in the constants $Z$; for internal consistency
we suppose that $g_n \simeq g_2^{n-2}$.

Let us represent the Taylor expansion of the function $V(h)$ as follows:
    \begin{equation}
V(h)=\sum^{\infty}_{n=2} \lambda_{n}h^{n}(x)/n!, \quad
V_{R}(h)=\sum^{\infty}_{n=2} Z_{n} \lambda_{n}h^{n}(x)/n!,
    \label{I14}
    \end{equation}
 In the following, we interpret similar objects as functions of a single variable $h(x)$, and $V'$, $V''$, etc., 
as the corresponding derivatives with respect to this variable. In this notation the matrix $W$ 
(under condition that $Z=1$) can be symbolically represented as
\begin{eqnarray}
W=\pmatrix{-\partial_{\parallel}^{2}h'\cdot V''& L^{T} \cr L&-2 \cr} \, ,
\label{I15}
\end{eqnarray}
where 
$L\equiv \partial_{t}-\nu_{\parallel}\partial_{\parallel}^{2}-\nu_{\bot}\partial_{\bot}^{2}-\partial_{\parallel}^{2}V'$,
and $L^T\equiv -\partial_{t}-\nu_{\parallel}\partial_{\parallel}^{2}-\nu_{\bot}
\partial_{\bot}^{2}-V'\partial_{\parallel}^{2}$ is the transposed operation.

In order to calculate the constants $Z$ we need only the divergent part of expression (\ref{W19}), which was 
previously established to have the form   
\[ \int\ dx\partial^{2}h'(x) R(h(x)) \]
with a function $R(h)$ similar to $V(h)$. This means that we need to calculate Tr ln
in (\ref{W20}) with matrix (\ref{I15}) only to the first order in its 
$hh$-element $-\partial_{\parallel}^{2}h'\cdot V''$. 
We can do this employing the well-known formula $\delta({\rm Tr\ ln}K)={\rm Tr}(K^{-1}\delta K)$
for any variation $\delta K$. By 
varying only the $hh$-element of the matrix $W$ we obtain 
\begin{eqnarray}
\int\ dx\partial^{2}h'(x) R(h(x)) \simeq -{\rm Tr}\,[D_{hh}V''\partial_{\parallel}^{2}h'] =
\nonumber \\ =
-\int dx\ D^{(hh)}(x,x) V''(h(x)) \partial_{\parallel}^{2}h'(x),
\label{I16}
\end{eqnarray}
where $D^{hh}=(W^{-1})_{hh}$ at $h'=0$. By the definition, $D^{hh}$ is the ordinary propagator 
$\langle hh \rangle$ of the model (\ref{RenAct}) with $Z=1$ and with 
$\nu_{\parallel}\partial_{\parallel}^{2}+\nu_{\bot}\partial_{\bot}^{2}+\partial_{\parallel}^{2}V'$ 
substituted for $\nu_{\parallel}\partial_{\parallel}^{2}+\nu_{\bot}\partial_{\bot}^{2}$.

There is another consideration that must be taken into account. After $\partial^2_{\parallel}$ is 
moved to the external factor $h'$ only a logarithmically divergent expression remains in the counterterm. 
This means that we can set all its external momenta to zero while calculating the divergent part of a given 
diagram (IR regularization is ensured by the cutoff). In its turn, this leads to the fact that we can ignore 
the inhomogeneity of $\partial^2_{\parallel}h'(x)$ and $h(x)$ (both can be assumed to be constant) in (\ref{I16}) 
when we select the poles in $\varepsilon$. Then $D_{hh}(x,x)$ can easily be calculated by going over to the 
momentum-frequency representation: 

\begin{eqnarray}
D_{hh}(x,x)=\int\int \frac{d\omega d{\bf k}}{(2\pi)^{d+1}} \,
\frac{2}{\omega^{2}+[\nu_{\parallel}k_{\parallel}^{2}+\nu_{\bot}k_{\bot}^{2}+k_{\parallel}^{2}V']^{2}} =
\nonumber \\ =
\frac{S_d}{(2\pi)^d}\frac{\mu^{-\varepsilon}}{ \varepsilon}\frac{1}{\sqrt{\nu_{\bot}(\nu_{\parallel}+V')}} +\dots ,
\label{I17}
\end{eqnarray}
where the elipsis stands for the UV-finite part. 

Substituting (\ref{I16}) and (\ref{I17}) into (\ref{W20}) yields the following expression for the divergent part of  
$\Gamma_{1}(\Phi)$ with the required accuracy:
\begin{equation}
\Gamma_{1}(\Phi)= \frac{S_d}{2(2\pi)^d}\frac{\mu^{-\varepsilon}}{ \varepsilon}\int dx
\frac{V''(h(x))}{ \sqrt{ \nu_{\bot}(\nu_{\parallel}+V'(h(x)))} }\, \partial^{2}h'(x)
\label{I18}
\end{equation}

We can find the one-loop contributions of order $1/\varepsilon$ in all constants $Z$ due to the fact that the 
sum of (\ref{I18}) and the loopless contribution in (\ref{W20}) has no pole in $\varepsilon$ (it cancels out).

Let us introduce the representation
\begin{equation}
\frac{V''(h(x))}{\sqrt{ \nu_{\bot}(\nu_{\parallel}+V'(h(x)))}} =\sum^{\infty}_{n=0}
\mu^{\varepsilon(n+1)/2} \nu_{\bot}^{(n-1)/4} \nu_{\parallel}^{(n+3)/4}\frac{r_{n}h^{n}}{n!},
\label{I19}
\end{equation}
for the Taylor expansion of the integrand in (\ref{I18}).

Then $r_{n}$ are completely dimensionless coefficients -- polynomials in the charges $g_n$. 
Combining the above condition for the canceling out of poles in $\varepsilon$ and (\ref{I6}), we get   
\begin{equation}
Z_{\bot}=1, \quad
Z_{\parallel}=1-\frac{r_{1}S_d}{2(2\pi)^d\varepsilon}+\dots\,\\
Z_{n}=1-\frac{r_{n}}{g_{n}}\frac{S_d}{2(2\pi)^d\varepsilon}+\dots\, .
\label{I20}
\end{equation}

The operation ${\stackrel{\sim}{\cal D}}_{\mu}$ in (\ref{betagw})
assumes the form
\[ {\stackrel{\sim}{\cal D}}_{\mu} = \sum_{n}
\left({\stackrel{\sim}{\cal D}}_{\mu}
g_{n}\right)\partial_{g_{n}} = \sum_{n} \beta_{n}\partial_{g_{n}}. \]
So in order to achieve the required accuracy it is sufficient to use only the first terms 
in the $\beta$-functions (\ref{betagw}). 
This yields 

\begin{equation}
{\stackrel{\sim}{\cal D}}_{\mu} \simeq -\frac{\varepsilon}{2} {\cal D}_{g},
\quad
{\cal D}_{g}= \sum^{\infty}_{n=2} (n-1) g_{n} \partial_{g_{n}}.
\label{I21}
\end{equation}

This consideration together with (\ref{I20}), (\ref{I8}), and (\ref{betagw}) leads to the following expressions for 
the one-loop RG-functions:

\begin{equation}
\gamma_{\parallel}=a{\cal D}_{g} r_{1}/2, \quad a\equiv \frac{S_d}{2(2\pi)^d};
\label{I22a}
\end{equation}
\begin{equation}
\beta_{n}=-\varepsilon\frac{n-1}{2}g_{n}+\frac{n+3}{4}g_{n}\gamma_{\parallel}-\frac{a}{2}({\cal D}_{g}
-n+1) r_{n}.
\label{I22b}
\end{equation}

The explicit expressions for the first four coefficients $r_n$ [the first term with $r_0$ in (\ref{I19}) contributes nothing to (\ref{I18})] are found from the definitions (\ref{I19}), (\ref{I14}), (\ref{I6}):

\[ r_{1}=g_{3}-\frac{1}{2}g_{2}^{2},\quad r_{2}=g_{4}-\frac{3}{2}g_{2}g_{3}+\frac{3}{4}g_{2}^{3}, \]
\[ r_{3}=g_{5}-2g_{2}g_{4} -\frac{3}{2}g_{3}^{2}+\frac{9}{2}g_{2}^{2}g_{3}-\frac{15}{8}g_{2}^{4}, \]
\[  r_{4}=g_{6}-\frac{5}{2}g_{2}g_{5} +\frac{15}{2} g_{2}^{2} g_{4} -5 g_{3}g_{4}+
\frac{45}{4} g_{2}g_{3}^{2}-\frac{75}{4} g_{2}^{3}g_{3} +\frac{105}{16} g_{2}^{5} , \]

when substituted into (\ref{I22a}) they yield: 

\begin{eqnarray}
\gamma_{\parallel} &=& \frac{a}{2} (2g_{3}-g_{2}^{2}),
\label{I23a} \\
\beta_{2}&=&-\frac{\varepsilon}{2} g_{2}+a(-g_{4} +\frac{11}{4} g_{2}g_{3}-\frac{1}{8} g_{2}^{3}),
\nonumber \\
\beta_{3}&=&-\varepsilon g_{3}+a(-g_{5}+2g_{2}g_{4}+3 g_{3}^{2}-\frac{21}{4} 
g_{2}^{2}g_{3} +\frac{15}{8}  g_{2}^{4})
\label{I23b}
\end{eqnarray}

We recall that we have to admit $g_{n} \sim g_{2}^{(n-1)}$ for the sake of consistency of the approximation.

\section{Attractors and critical exponents} \label{sec:Att}

Let us turn to the complete system (\ref{I23b}) of the $\beta$-functions. The fixed points of RG equations 
 can be found from the requirement that  $\beta_{n}(g_{*})=0$, $n=2,3,\dots$. 
The explicit form of the $\beta$-functions (\ref{I23b}) shows that we can choose the coordinates 
$g_{2*}$, and $g_{3*}$ arbitrarily, while all the other $g_{n*}$ with $n\ge4$ are then  uniquely determined 
from the equations $\beta_{k}(g_{*})=0$, $k\ge3$. This means that in the infinite-dimensional space of the 
couplings $g\equiv \{ g_{n} \}$ the RG-equation (\ref{RG2}) has  a two-dimensional surface of fixed points, 
parametrized by the values of $g_{2*}$, and $g_{3*}$. 

In general case, studying these points is a difficult task. However, according to the general rule 
\cite{Zinn}, a point $g_{*}$ is IR stable if the real parts of all the eigen-numbers of the matrix 
$\omega_{nm}=\partial\beta_{n}/\partial g_{m}|_{g_{*}} $ are strictly positive. The requirement that  
all the diagonal elements $\omega_{nn}$ be positive is the necessary  condition for IR-stabihty. 
Equation (\ref{I22b}) can be used to calculate these elements for all values of $n$: 

\[ \omega_{22}=-\frac{\varepsilon}{2}+a\left[\frac{11}{4} g_{3*}-\frac{3}{8} g_{2*}^{2}\right],\quad
\omega_{33}=-\varepsilon+a \left[6g_{3*}-\frac{21}{4} g_{2*}^{2}\right], \]
and for $n\ge4$ we have
\[ \omega_{nn}= -\varepsilon\frac{n-1}{2}+a \frac{(n+1)^2+2}{4} g_{3*}-\frac{a}{8} (n(3n+4)+3)g_{2*}^{2}. \]

In  a certain  region $g_{3*}\ge 7g_{2*}^{2}/8+ \varepsilon/6$ all these quantities are positive. 
Of course, this is just a necessary condition; still, we can assume that the surface of fixed points 
$g_{*}$ contains a region of IR stability. If this is indeed so,  the model may contain IR scaling with 
nonuniversal critical dimensions ({\it i.e.} there is a dependence on the the parameters  $g_{2*}$, and $g_{3*}$). 

In dynamic models of the type (\ref{act1}) the critical exponents $\Delta_F$ of an arbitrary quantity
 $F$ (a field or a parameter) is given by the following expression:
\begin{eqnarray}
\Delta_{F} = d^{\bot}_{F} + d^{\parallel}_{F}\Delta_{\parallel} +d^{\omega}_{F}\Delta_{\omega}  +
\gamma_{F}^{*}, \quad \Delta_{w} =2-\gamma_{\bot}^{*}, \quad \Delta_{\parallel}= 1 + \gamma_{\parallel}^{*}/2.
\label{Dimension}
\end{eqnarray}

In case at hand for $F= h$ we have $\gamma_{h}^{*}=0$ and $\gamma_{\bot}^{*}=0$ (the fields and the parameter 
$\nu_{\bot}$ are not renormalized). Relations (\ref{Dimension}) together with the table~1 yield the exact result 
$2\Delta_{h}=d-1+\Delta_{\parallel}-\Delta_{\omega}$; from (\ref{I23a}) we find in the one-loop approximation 
that $\Delta_{\parallel}=1+a (2g_{3*}-g_{2*}^{2})/4$, $\Delta_{h}=a (2g_{3*}-g_{2*}^{2})/8$. 

\section{Conclusion} \label{sec:Conc}

We applied to the modified model \cite{Pastor1,Pastor2} the standard field
theoretic RG. It turned out that the model can be reformulated as a renormalizable field theoretic model
with an infinite set of independent renormalization constants (thus, infinite set of coupling constants). 
Indeed, to construct renormalizable model it is necessarily to include infinitely many coupling
constants, and the corresponding RG equations involve infinitely many
$\beta$-functions. Despite this fact, it appears possible to derive the one-loop counterterm employing the 
method, earlier proposed in  \cite{AV} for  an isotropic model
of surface roughening. The method yields a two-dimensional surface of fixed points which is likely to 
contain IR attractive region(s). Indeed, experimental results (see the discussion in~\cite{Pastor2}) 
indicate two wide ranges of roughening exponent value which might be explained by the existence of two different 
IR attractive regions. 

As the model needs to contain infinite set of coupling constants to be renormalizable it seems that truncated models 
like ~\cite{Pastor1,Pastor2} or the one with odd $V(h)$ might not be suitable for the RG analysis. 
The naive approach of putting the corresponding coupling constants in $V(h)$ to zero in  attempt to compare the 
results shows that in the case of the model ~\cite{Pastor1,Pastor2} there is no agreement. 

To compare the critical exponents of those two models one has to identify
$z_{\bot}=\Delta_{\omega}$, $\zeta_{\bot}=\Delta_{\parallel}$,
$\alpha_{\bot} =\Delta_{h}$. Obvious calculations show that in the case of $\lambda_n=0$ for all $n$ but $n=3$ 
the critical exponents are $\Delta_{\omega}=2$, $\Delta_{\parallel}=1+(2-d)/6$, and
$\Delta_{h}=(2-d)/12$. The last two values differ from the ones reported in~\cite{Pastor1,Pastor2}.

For odd $V(h)$ a two-dimensional surface of fixed points reduces to a curve. From the symmetry considerations, 
as well as from the explicit expression for the coumtertem (\ref{I18}),
it is clear that this case is renormalizable in itself. One can simply set all the odd couplings $g_{2n+1}$ and
the corresponding $\beta$ functions in (\ref{I22b}) equal to zero.

If the surface of fixed points does indeed contain IR attractive regions, than the model exhibits scaling behaviour. 
The corresponding scaling exponents turn out to be nonuniversal because of their dependence
on the coordinates of specific fixed point on the surface (curve). Nonetheless, they satisfy certain exact relations.

In the further study, it would be interesting to investigate how the model behaves if there is a turbulent 
velocity field involved; for the isotropic case, see \cite{AA}.

From a more theoretical point of view, it is desirable to write down the RG equations and to find the fixed point(s) 
directly in terms of the function $V(h)$, so that instead of infinitely many $\beta$ functions for the infinite set
of couplings $g_n$ we would have the only $\beta(V)$ functional with the only functional argument $V(h)$; 
see the discussion in \cite{Dima} 
for a general case.

This work remains for the future and is partly in progress. 

\section*{Acknowledgments}
The authors thank L. Ts. Adzhemyan, M. Hnatich, J. Honkonen and M. Yu. Nalimov for discussion.
The authors thank the Organizers of the International Conference
``Models in Quantum Field Theory V'' for the opportunity to present the
results of their research.
The authors also acknowledge the Saint-Petersburg State University for
research grant 11.38.185.2014. One of the authors (P.K.) was also supported by the RFBR research grant 16-32-00086.

\end{document}